\documentstyle[preprint,aps,epsfig,amsbsy,floats]{revtex}
 \tightenlines

\begin{document}
\newcommand{\beq}{\begin{equation}}
\newcommand{\eeq}{\end{equation}}
\newcommand{\beqa}{\begin{eqnarray}}
\newcommand{\eeqa}{\end{eqnarray}}
\newcommand{\sr}{\sqrt}
\newcommand{\fr}{\frac}
\newcommand{\mn}{\mu \nu}
\newcommand{\G}{\Gamma}

\draft \preprint{hep-th/0402198,~ INJE-TP-04-02}
\title{Second-order corrections to noncommutative spacetime  inflation}
\author{ Hungsoo Kim, Gil Sang Lee, Hyung Won Lee, and  Yun Soo Myung\footnote{E-mail address:
ysmyung@physics.inje.ac.kr}}
\address{
Relativity Research Center and School of Computer Aided Science\\
Inje University, Gimhae 621-749, Korea} \maketitle

\begin{abstract}
We investigate how the uncertainty of noncommutative spacetime
affects on inflation. For this purpose, the noncommutative
parameter $\mu_0$ is taken to be  a zeroth order slow-roll
parameter. We  calculate the noncommutative power spectrum up to
second order using the slow-roll expansion. We find corrections
arisen from a change of the  pivot scale
 and  the presence of a variable noncommutative parameter,
 when comparing with the commutative power spectrum.
The power-law inflation is chosen to obtain  explicit forms for
the power spectrum, spectral index, and running spectral index. In
cases of the power spectrum and spectral index, the noncommutative
effect of higher-order corrections  compensates for a loss of
higher-order corrections in the commutative case. However, for the
running spectral index, all higher-order corrections to the
commutative case always provide  negative spectral indexes, which
could explain the recent WMAP data.
\end{abstract}

\thispagestyle{empty}
\setcounter{page}{0}
\newpage
\setcounter{page}{1}

\section{Introduction}
String  theory as a candidate for the theory of everything can say
something  about cosmology\cite{Bran1}. Focusing on a universal
property of string theory, it is very interesting to study its
connection to cosmology. The universal property which we wish to
choose here is a new uncertainty relation of $ \triangle t_p
\triangle x_p \ge l^2_s$ where $l_{s}$ is the string length
scale\cite{Yone}. This implies that spacetime is noncommutative.
It is compared to a stringy uncertainty relation of  $ \triangle
x_p \triangle p \ge 1+ l^2_s \triangle p^2$. The former is
considered as a universal property for strings as well as
D-branes, whereas the latter is suitable only for strings.
Spacetime noncommutativity does not affect the evolution of the
homogeneous background. However, this leads to a coupling between
the fluctuations generated in inflation and the flat background of
Friedmann-Robertson-Walker (FRW) spacetime\cite{BH}. Usually the
coupling appears to be nonlocal in time.

On the other hand, it is generally accepted that curvature
perturbations produced during inflation are considered  to be
 the origin of  inhomogeneities necessary for
explaining  galaxy formation and other large-scale structure. The
first year results of WMAP put forward more constraints on
cosmological models and confirm the emerging standard model of
cosmology, a flat $\Lambda$-dominated universe seeded by
scale-invariant adiabatic gaussian fluctuations\cite{Wmap1}. In
other words, these results coincide with  predictions of the
inflationary scenario with an inflaton.  Also WMAP brings about
some new intriguing results: a running spectral index of scalar
metric perturbations and an anomalously low quadrupole of the CMB
power spectrum\cite{Wmap2}. If inflation is affected by physics at
a short distant close to string scale, one expects that the
spacetime uncertainty must be encoded in the CMB power
spectrum\cite{LMMP}. For example, the noncommutative power-law
inflation may produce a large running spectral index to fit the
data of WMAP\cite{HM1,TMB,HM2}.

Recently the noncommutative power spectrum, spectral index, and
running spectral index of the curvature perturbations produced
during inflation have been calculated  with  the slow-roll
parameters $\epsilon_1$ and $\delta_n$ and noncommutative
parameter $\mu_0$\cite{HM3}. Authors in\cite{KLM}  have examined
whether or not the noncommutative parameter is considered as a
slow-roll parameter. It turned out that the noncommutative
parameter $\mu_0$ is considered as a zeroth order slow-roll
parameter.  In this work, we will make a further progress in this
direction. We calculate the noncommutative power spectrum up to
second order in the slow-roll expansion and up to first order in
the noncommutative parameter. Also the spectral index and running
spectral index are obtained for the noncommutative spacetime
inflation. By choosing the power-law inflation, we find
corrections to the commutative inflation. These result from a
change of the pivot scale and the presence of $\mu_0
\not=$constant.  In order to understand the role of $\mu_0$
further, we also calculate the power spectrum using
$\mu_0$=constant, and the spectral index and running spectral
index using $\fr{d\mu_0}{d \ln k}\simeq-4\mu_0\epsilon_1$.

The organization of this work is as follows. In Sec. II we review
the slow-roll approximation in the noncommutative spacetime
inflation. We calculate the power spectrum up to second order
using the slow-roll expansion with $\mu_0 \neq$constant in Sec.
III. Sec. IV is devoted to obtaining the power spectrum up to
second-order corrections by making use of the slow-roll expansion
with $\mu_0=$ constant. This gives us an intermediate result
between the slow-roll approximation and slow-roll expansion.
Finally we discuss our results in Sec. V.

\section{Slow-roll approximation}

Our starting point is the effective
action during inflation,
\begin{equation}
S = \int \left[ -\frac{M^{2}_{Pl}}{2}R + \frac{1}{2}(\partial_\mu
\phi)^2 - V(\phi) \right] \sqrt{-g} \ d^4 x,
\end{equation}
where $M^{2}_{P}$ is the reduced Planck mass defined by
$M_{P}=(8\pi G)^{1/2}$. For simplicity we choose $M^{2}_{P}=1$.
 The scalar  metric perturbation to the homogeneous, isotropic
background  is expressed  in the longitudinal gauge as
\cite{bardeen}
\begin{equation}
\label{con-p}
 ds^2_{con-p} = a^2(\eta) \left\{(1+2A) d\eta^2
-(1+2\psi)d{\bf x}\cdot d{\bf x} \right\},
\end{equation}
where the conformal time $\eta$ is given by $d\eta = dt / a $. We
get a relation of $\psi=A$ because the stress-energy tensor does
not have any off-diagonal component. It is convenient to express
the density perturbation in terms of the curvature perturbation
$R_c$ of comoving hypersurfaces given by~\cite{lukash}
\begin{equation}
R_c = \psi-\frac{H}{ \dot \phi} \delta \phi
\end{equation}
during inflation, where $\delta\phi$ is the perturbation in
inflaton: $\phi({\bf x},\eta)=\phi(\eta)+\delta\phi({\bf
x},\eta)$. The overdot is derivative with respect to a comoving
time $t$ defined in the flat FRW line element:
$ds^2_{FRW}=dt^2-a(t)^2 d{\bf x}\cdot d{\bf x}$. Introducing
\begin{equation}
z \equiv \frac{a\dot \phi}{H} \ \ \ \mbox{and}\ \ \ \varphi \equiv
a\left(\delta\phi-\frac{\dot\phi}{H} \psi\right) = -z R_c,
\end{equation}
the bilinear action for curvature  perturbation is \cite{mukhanov}
\begin{equation}
\label{bilinear} S = \int \frac{1}{2} \left[
\left(\frac{\partial\varphi}{\partial\eta}\right)^2 -
\left(\nabla\varphi\right)^2 +
\left(\frac{1}{z}\frac{d^2z}{d\eta^2}\right)\varphi^2 \right]
d\eta\,d^3{\bf x}.
\end{equation}
At this stage we wish to note that $z$ encodes information about
inflation.  Because the background is spatially flat, we can
expand all perturbed fields in terms of Fourier modes as $
\varphi({\bf x},\eta)=\int \fr{d^3{\bf k}}{(2 \pi)^{3/2}}
\varphi_{{\bf k}}(\eta) e^{i{\bf k}\cdot {\bf x}}$. In second
quantization, these modes are given by $\varphi_{{\bf
k}}(\eta)=b({\bf k})\varphi_k(\eta) +b^\dagger(-{\bf
k})\varphi^*_k(\eta)$. This quantum-to-classical behavior is a
great success for the theory\cite{LL}. If it had failed,
prediction for the power spectrum  would have had nothing to do
with reality.

For convenience we introduce another time coordinate $\tau$ to
incorporate the noncommutative spacetime. Then the perturbed
metric in Eq.(\ref{con-p}) can be rewritten as
\begin{equation}
ds^2_{non-p} = a^{-2}(\tau)(1+2A) d\tau^2 - a^2(\tau)(1+2\psi)
d{\bf x}\cdot d{\bf x}.
\end{equation}
The spacetime uncertainty relation of $ \triangle t_p \triangle
x_p \ge l^2_s$ becomes
\begin{equation}\label{non-c}
\triangle \tau \triangle x \ge l^2_s \eeq for a cosmological
purpose. This stringy spacetime uncertainty relation is compatible
with a FRW background  spacetime. However a nonlocal coupling
between the fluctuation mode and the background arises because of
the $*$-product.  We propose the transition to noncommutative
spacetime obeying Eq.(\ref{non-c}) by taking the operator
appearing in the bilinear action in Eq.(\ref{bilinear}) and
replacing all multiplications by $*$-products\cite{BH}.  Using the
Fourier transform, the uncertainty relation leads to the modified
action for $\tilde{\varphi}_k$: \beq \label{modact}\tilde{S}
=\fr{V_T}{2} \int d\tilde{\eta} ~d^3{\bf
k}~z_k^2(\tilde{\eta})~\Big[\fr{d\tilde{R}_{c-{\bf
k}}}{d\tilde{\eta}}\fr{d\tilde{R}_{c{\bf k}}}{d\tilde{\eta}}-
k^2\tilde{R}_{c-{\bf k}}\tilde{R}_{c{\bf k}}\Big], \eeq where
$V_T$ is the total spatial volume, and $\tilde{\eta}$ is a new
conformal time which is suitable for a noncommutative spacetime
infation. Here $z_k$ is some smeared version of $z$ or $a$ over a
range of time of characteristic scale $\triangle \tau =l_s^2k$
defined by \beq \label{zzz}
z^2_k(\tilde{\eta})=z^2y^2_k(\tilde{\eta}),
~~y^2_k=\sqrt{\beta^+_k\beta^-_k},~~\fr{d\tilde{\eta}}{d\tau}
=\sqrt{\fr{\beta^-_k}{\beta^+_k}}, \eeq where \beq \label{betak}
\beta^{\pm}_k=\fr{1}{2} \left[a^{\pm 2}(\tau+l^2_sk)+a^{\pm
2}(\tau-l^2_sk)\right]. \eeq   This representation has the
advantage of preserving both spatial translational and rotational
symmetry of the flat FRW metric, in compared to constructions
based on the conventional noncommutative relations: $
\left[x^\mu,x^\nu\right]=i \theta^{\mn}$\cite{BG,FKM}. Actually a
spacetime noncommutativity does not affect the evolution of the
homogeneous background. However, this leads to a coupling between
the fluctuations generated in inflation and the flat background of
FRW space. The coupling appears to be nonlocal in time as is shown
in Eq.(\ref{betak}). Actually $\triangle \tau=l_s^2k$ in
Eq.(\ref{non-c}) induces  the uncertainty of  time in defining
$a$. If one does not require the uncertainty relation, one finds
easily commutative relations that $y_k \to 1, ~z_k \to z,~
\tilde{\eta}\to \eta,~\tilde{\varphi}_k(\tilde{\eta}) \to
\varphi_k(\eta)$.

From Eq.(\ref{modact}), we find
 the Mukhanov-type equation
\begin{equation}
\label{eqsn} \frac{d^2\tilde{\varphi}_k}{d\tilde{\eta}^2} +
\left(k^2-\frac{1}{z_k}\frac{d^2z_k}{d\tilde{\eta}^2}\right)\tilde{\varphi}_k
= 0.
\end{equation}
Our task is to solve Eq.~(\ref{eqsn}). For this purpose we
introduce slow-roll parameters $\mu_0,~\epsilon_1,~\delta_n$
defined as
\begin{equation}
\mu_0=\Big(\fr{kH}{aM^2_s}\Big)^2,~~\epsilon_1 = -\frac{\dot
H}{H^2} =
\frac{1}{2}\left(\frac{\dot{\phi}}{H}\right)^2,~~\delta_n \equiv
\frac{1}{H^n\dot{\phi}}\frac{d^{n+1}\phi}{dt^{n+1}}.
\end{equation}
 Here
$M_s=1/l_s$ and  the subscript denotes the order in the slow-roll
expansion.
 We use the slow-roll approximation which means
that   $\epsilon_1,~\delta_1,~\mu_0$ are taken to be approximately
constant in calculation of the noncommutative power spectrum. That
is, $\epsilon_1,~\delta_1,~\mu_0=$constant.  For this purpose, we
obtain relations up to first order\cite{HM3}
 \beq \label{nonzzz}
\frac{1}{z_k}\frac{d^2z_k}{d\tilde{\eta}^2}\simeq
2(aH)^2\Big(1+\epsilon_1+\fr{3}{2}\delta_1 -2\mu_0\Big) \eeq and
\beq \label{nonaH}aH\simeq
-\fr{1}{\tilde{\eta}}(1+\epsilon_1+\mu_0).
 \eeq
 Then Eq.(\ref{eqsn}) takes the form
 \beq
\frac{d^2\tilde{\varphi}_k}{d\tilde{\eta}^2} +
\left(k^2-\frac{(\nu^2-\fr{1}{4})}{\tilde{\eta}^2}\right)\tilde{\varphi}_k
= 0\eeq with $\nu=\fr{3}{2}+2\epsilon_1+\delta_1$. We note here
that this equation takes the same form as in the commutative
case\cite{SL}. The asymptotic solution to Eq.(\ref{eqsn}) in the
limit of $-k\tilde{\eta} \to \infty$ takes the form
\begin{equation}\label{nbc}
\tilde{\varphi}_k =\frac{1}{\sqrt{2k}}e^{-ik\tilde{\eta}}.
\end{equation}
In the limit of $-k\tilde{\eta} \to 0$, one finds asymptotic form
of the Hankel function $H^{(1)}_\nu(-k\tilde{\eta})$ \beq
\label{nonphi}\tilde{\varphi}_k \simeq
e^{i(\nu-\fr{1}{2})}2^{\nu-\fr{3}{2}}\fr{\Gamma(\nu)}{\Gamma(\fr{3}{2})}
\fr{1}{\sqrt{2k}}(-k\tilde{\eta})^{\fr{1}{2}-\nu}. \eeq
Furthermore, from Eqs.(\ref{zzz}) and (\ref{betak}), we have an
expression up to first order\beq \label{yk}y_k\simeq 1+\mu_0. \eeq
The Fourier transform of curvature perturbation is given by
$\tilde{R}_k=-\tilde{\varphi}_k(\tilde{\eta})/z_k$.
 Then the  noncommutative power spectrum is defined  by
\begin{equation}\label{nps}
\widetilde{P}_{R_c}(k) = \left(\frac{k^3}{2\pi^2}\right)
\lim_{-k\tilde{\eta}\rightarrow0}\left|\frac{\tilde{\varphi}_k}{z_k}\right|^2.
\end{equation}
Substituting equations (\ref{nonaH}), (\ref{nonphi}) and
(\ref{yk}) into Eq.(\ref{nps}), one finds
 \beq
\widetilde{P}_{R_c}(k)
=\frac{H^4}{(2\pi)^2\dot\phi^2}\Big[2^{\nu-\fr{3}{2}}\fr{\Gamma(\nu)}{\Gamma(\fr{3}{2})}\Big]^2
\Big(\fr{k}{aH}\Big)^{-2(2\epsilon_1+\delta_1)}
\fr{1}{(1+\epsilon_1+\mu_0)^{2(1+2\epsilon_1+\delta_1)}(1+\mu_0)^{2}}.
\eeq Making use of  the Taylor expansions up to first order as\beq
2^{\nu-\fr{3}{2}}\fr{\Gamma(\nu-\fr{3}{2}+\fr{3}{2})}{\Gamma(\fr{3}{2})}
\simeq
1+\alpha(2\epsilon_1+\delta_1),~~e^{-2(2\epsilon_1+\delta_1)\ln\Big(\fr{k}{aH}\Big)}\simeq
1- 2(2\epsilon_1+\delta_1)\ln\Big(\fr{k}{aH}\Big),\eeq we have
\cite{KLM}
 \beq \label{nonps1} \tilde{P}^{1st}_{R_c}(k)
= \frac{H^4}{(2\pi)^2\dot\phi^2} \left\{ 1 -2\epsilon_1 -4\mu_0 +2
\left(\alpha-\ln\Big(\fr{k}{aH}\Big)\right)(2\epsilon_1+\delta_1)
\right\}.  \eeq  In the limit of $\mu_0 \to 0$,
$\widetilde{P}^{1st}_{R_c}(k)$ reduces to the commutative power
spectrum\cite{MS1,STW}. In the noncommutative spacetime approach
the horizon crossing occurs at
$k^2=\frac{1}{z_k}\frac{d^2z_k}{d\tilde{\eta}^2}$\cite{BH}. Hence,
from, Eq.(\ref{nonzzz}) we use the other pivot scale $k_{*}=
\sqrt{2}aH$. As a result, we obtain the noncommutative power
spectrum up to first order as
 \beq \label{nonps11}\tilde{P}^{1st}_{R_c}(k)
=\left. \fr{1}{2\epsilon_1}\frac{H^2}{(2\pi)^2} \left\{ 1
-2\epsilon_1 -4\mu_0+ 2\alpha_*(2\epsilon_1+\delta_1)
\right\}\right|_{k=\sqrt{2}aH} \eeq with
$\alpha_*=\alpha-\fr{\ln2}{2}$. Here we note that the right hand
side is evaluated at $k=\sqrt{2}aH$.

 Let us compare Eq.(\ref{nonps11})  with the
commutative power spectrum.  An apparent change of the  pivot
scale from $k_c=aH$ to $k_{*}=\sqrt{2}aH$ amounts to replacing
$\alpha=0.7296$ by $\alpha_*=0.3831$ in the first-order
calculation\cite{STW}. Further we note that
$k_c=a(\tilde{\eta}_c)H(\tilde{\eta}_c)$ and
$k_*=\sqrt{2}a(\tilde{\eta}_*)H(\tilde{\eta}_*)$. This means that
the noncommutative horizon crossing time
$\tilde{\eta}=\tilde{\eta}_*$ is different from
$\tilde{\eta}=\tilde{\eta}_c$. In order to make its correction, we
have to establish a connection between $H(\tilde{\eta}_c)$ and
$H(\tilde{\eta}_*)$, $\epsilon_1(\tilde{\eta}_c)$ and
$\epsilon_1(\tilde{\eta}_*)$, and so on\cite{MS2}. At the level of
slow-roll approximation, we consider slow-roll parameters to be
constant everywhere, except in the factor $1/\epsilon_1$.
Performing a series expansion around $\tilde{\eta}=\tilde{\eta}_c$
leads to \beq
\fr{1}{\epsilon_1(\tilde{\eta}_*)}=\fr{1}{\epsilon_1(\tilde{\eta}_c)}
[1-2(\epsilon_1(\tilde{\eta}_c)+\delta_1(\tilde{\eta}_c))
\triangle \tilde{\eta}+
\cdots],~H^2(\tilde{\eta}_*)=H^2(\tilde{\eta}_c)[1-2\epsilon_1(\tilde{\eta}_c)\triangle
\tilde{\eta}\cdots] \eeq and \beq
\ln\Big[\fr{k}{a(\tilde{\eta}_*)H(\tilde{\eta}_*)}\Big]=
\ln\Big[\fr{k}{a(\tilde{\eta}_c)H(\tilde{\eta}_c)}\Big]-\triangle
\tilde{\eta}+\cdots, \eeq where the dots represent first-order
terms and $\triangle \tilde{\eta}=\tilde{\eta}_*-\tilde{\eta}_c$.
Substituting everything into Eq.(\ref{nonps1}), one finds that all
$\triangle \tilde{\eta}$ cancel. When $\mu_0=0$, the result is the
same form as in the commutative case except replacing $\alpha$ by
$\alpha_*$
 \beq \label{ps11}\tilde{P}^{1st}_{R_c}(k)
=\left. \fr{1}{2\epsilon_1}\frac{H^2}{(2\pi)^2} \left\{ 1
-2\epsilon_1 -4\mu_0+ 2\alpha_*(2\epsilon_1+\delta_1)
\right\}\right|_{k=aH}.\eeq Hence we could use the slow-roll
expansion at $k=k_c$ to obtain the noncommutative power spectrum
at $k=k_*$. As is shown in Eq.(\ref{ps11}), the noncommutative
effect is to reduce the power spectrum by making use of $\alpha
\to \alpha_*$ and $\mu_0\not=0$.

\section{Slow-roll expansion with $\mu_0\not=$ constant}

The slow-roll approximation could not be considered as a correct
approach to calculate the power spectrum even for up to first
order.  In order to calculate the power spectrum up to second
order correctly, one should use the slow-roll expansion based on
Green's function technique. In the case of $\mu_0$=0, two of the
slow-roll approximation and slow-roll expansion give the same
power spectrum up to first order. However, in the case  of $\mu_0
\not=0$, two provide  different results. The assumption that
slow-roll parameters are taken to be constant in the slow-roll
approximation is not generally true.

The key step is to introduce a variable nature of slow-roll
parameters during inflation: \beq \label{slow-di}\dot \mu_0=-4
H\mu_0\epsilon_1,~~\dot \epsilon_1
=2H(\epsilon_1^2+\epsilon_1\delta_1),~~\dot{\delta}_1=
H(\epsilon_1\delta_1-\delta^2_1+\delta_2) \eeq which means that
the derivative of slow-roll parameters with respect to time
increases their  order by one in the slow-roll expansion.
 However this expansion is useful for deriving the power
spectrum at $k=aH$ but not $k=\sqrt{2}aH$ and thus it works well
for commutative case and higher order case. As far as we know,
there is no way to calculate the power spectrum up to second order
at an arbitrary pivot scale using the slow-roll expansion. Hence,
first we use the slow-roll expansion at $k=aH$ to calculate the
noncommutative power spectrum up to second order. Then, assuming
the rule that the change of pivot scale from $k=aH$ to
$k=\sqrt{2}aH$ amounts to replacing $\alpha$ by $\alpha_*$, our
calculation may  provide the result up to second order that will
be derived from the slow-roll expansion at $k=\sqrt{2}aH$.

Using notations of $y = \sqrt{2k}\, \tilde{\varphi}_k$ and
$\tilde{x} = -k\tilde{\eta}$, we can reexpress Eq.(\ref{eqsn}) as
\begin{equation} \label{eqsnn}
\frac{d^2y}{d\tilde{x}^2} +
\left(1-\frac{1}{z_k}\frac{d^2z_k}{d\tilde{x}^2}\right)y=0.
\end{equation}In general its asymptotic solutions
are given by
\begin{equation}
\label{z-sol} y \longrightarrow \left\{
\begin{array}{l l l}
e^{i\tilde{x}} & \mbox{as} & \tilde{x} \rightarrow \infty \\ \\
\sqrt{2k}\,\tilde{A}_k z_k & \mbox{as} & \tilde{x} \rightarrow 0.
\end{array}
\right.
\end{equation}
We solve Eq.~(\ref{eqsnn}) with the boundary condition
Eq.~(\ref{z-sol}) to eventually calculate $\tilde{A}_k$. Now we
can choose the ansatz that $z_k$ takes the form
\begin{equation}
z_k = \frac{1}{\tilde{x}}\widetilde{f}(\ln \tilde{x}).
\end{equation}
Then we have
\begin{equation}
\frac{1}{z_k}\frac{d^2z_k}{d\tilde{x}^2} = \frac{2}{\tilde{x}^2} +
\frac{1}{\tilde{x}^2}\widetilde{g}(\ln \tilde{x}),
\end{equation}
where
\begin{equation}\label{gf}
\widetilde{g}=\frac{-3\widetilde{f}'+\widetilde{f}''}{\widetilde{f}}
\end{equation}
and the equation of motion is
\begin{equation}
\label{exactsol} \frac{d^2y}{d\tilde{x}^2} +
\left(1-\frac{2}{\tilde{x}^2}\right) y =
\frac{1}{\tilde{x}^2}\widetilde{g}(\ln \tilde{x}) y.
\end{equation}
The homogeneous solution with correct asymptotic behavior at
$\tilde{x} \rightarrow\infty$ is
\begin{equation}
\label{0sol} y_0(\tilde{x}) = \left(1 +
\frac{i}{\tilde{x}}\right)e^{i\tilde{x}}.
\end{equation}
Using Green's function technique, Eq.(\ref{exactsol}) with the
boundary condition Eq.(\ref{z-sol}) can be written as the integral
equation
\begin{equation}
\label{sol} y(\tilde{x}) = y_0(\tilde{x}) +
\frac{i}{2}\int_{\tilde{x}}^{\infty}du \ \frac{1}{u^2} \
\widetilde{g}(\ln u) \ y(u)
\left[y_0^*(u)y_0(\tilde{x})-y_0^*(\tilde{x})y_0(u)\right].
\end{equation}

We are now in a position to solve Eq.~(\ref{sol}) perturbatively
using the slow-roll expansion. Introducing
\begin{equation}\label{zexp}
\tilde{x}z_k = \widetilde{f}(\ln \tilde{x}) = \sum_{n=0}^{\infty}
\frac{\widetilde{f}_n}{n!}(\ln \tilde{x})^n,
\end{equation}
 $\widetilde{f}_n/\widetilde{f}_0$ is of order $n$ in the slow-roll expansion. This
expansion is useful for $\exp(-1/\xi) \ll \tilde{x} \ll
\exp(1/\xi)$ and for extracting information at $\tilde{x}=1$.

Considering a relation up to second order in the slow-roll
expansion and up to first order in $\mu_0$ as
\begin{equation}
\tilde{x} = -k\tilde{\eta} = -k\int d\tau
\left(\fr{\beta^-_k}{\beta^+_k}\right)^{1/2} \simeq
\frac{k}{aH}\left\{1+
\epsilon_1+3\epsilon_1^2+2\delta_1\epsilon_1+\mu_0(1-2\epsilon_1)
 \right\},
\end{equation}
we can express the expansion coefficients $\tilde{f}_n$ in terms
of $\epsilon_1$, $\delta_n$, and $\mu_0$ evaluated at $k=aH$. In
deriving the above expression,  we use Eq.(\ref{slow-di}). From
Eq.(\ref{zexp}) we  obtain
\begin{eqnarray}
\tilde{f}_2 &\simeq & \left.\frac{k\dot\phi}{H^2}\left\{
                               8\epsilon_1^2+9\epsilon_1\delta_1+\delta_2
                               +\mu_0(2\delta_2-14\epsilon_1\delta_1)
                           \right\}
\right|_{k=aH}, \\
\tilde{f}_1  &\simeq & \left.-\frac{k\dot\phi}{H^2}\left\{
                                2\epsilon_1+\delta_1+6\epsilon_1^2
                                +4\epsilon_1\delta_1+\mu_0(4\epsilon_1
                                -2\delta_1+14\epsilon_1^2+\epsilon_1\delta_1)
                            \right\}
\right|_{k=aH}, \\
\tilde{f}_0  &\simeq & \left. \frac{k\dot\phi}{H^2}\left\{
                         1+\epsilon_1+5\epsilon_1^2+3\epsilon_1\delta_1
                         +\mu_0(2+\epsilon_1+\delta_1+\epsilon_1^2+6\epsilon_1\delta_1)
                     \right\}
\right|_{k=aH},\\
\label{1/f0}\frac{1}{\tilde{f}_0}&\simeq &
\left.\frac{H^2}{k\dot\phi}
                            \left\{
                                1-\epsilon_1-4\epsilon_1^2-3\epsilon_1\delta_1
                                -\mu_0(2-3\epsilon_1+\delta_1-15\epsilon_1^2-8\epsilon_1\delta_1)
                            \right\}
                            \right|_{k=aH},\\
\label{f1/f0}\frac{\tilde{f}_1}{\tilde{f}_0}&\simeq & \left.
                           \left\{
                           -2\epsilon_1-\delta_1-4\epsilon_1^2-3\epsilon_1\delta_1
                           +\mu_0(8\epsilon_1+16\epsilon_1^2+10\epsilon_1\delta_1+\delta_1^2)
                           \right\}
                                     \right|_{k=aH},\\
\label{f2/f0}\frac{\tilde{f}_2}{\tilde{f}_0}&\simeq & \left.
                    \left\{
                        8\epsilon_1^2+9\epsilon_1\delta_1+\delta_2
                        -16\mu_0\epsilon_1(\epsilon_1+2\delta_1)
                    \right\}
\right|_{k=aH}.
\end{eqnarray}
Further Eqs.(\ref{gf}) and~(\ref{zexp}) give
\begin{equation}\label{gexp}
\tilde{g}(\ln \tilde{x}) = \sum_{n=0}^{\infty}
\frac{\tilde{g}_{n+1}}{n!}(\ln \tilde{x})^n,
\end{equation}
where $\tilde{g}_n$ is of order $n$ in the slow-roll expansion
and, up to second order \beq\label{g1} \tilde{g}_1  \simeq
-3\frac{\tilde{f}_1}{\tilde{f}_0}+\fr{\tilde{f}_2}{\tilde{f}_0},~~
\tilde{g}_2  \simeq
-3\Big(\frac{\tilde{f}_1}{\tilde{f}_0}\Big)^2-3\fr{\tilde{f}_2}{\tilde{f}_0}.
\eeq Expanding $y$ as
\begin{equation}\label{yexp}
y(\tilde{x}) = \sum_{n=0}^{\infty}y_n\left(\tilde{x}\right),
\end{equation}
where $y_0(\tilde{x})$ is the homogeneous solution in
Eq.(\ref{0sol}), and $y_n(\tilde{x})$ is of order $n$ in the
slow-roll expansion. Following the procedure in commutative
case~\cite{SG}, we solve Eq.~(\ref{sol}) perturbatively by
substituting Eqs.(\ref{gexp}) and~(\ref{yexp}) and equating terms
of the same order. We obtain the asymptotic form for $y$ up to
first-order corrections
\begin{eqnarray}\label{final-y}
y(\tilde{x}) & \rightarrow & i \left\{ \begin{array}{l}1 +
\frac{\tilde{g}_1}{3}
\left[\alpha+\frac{i\pi}{2}\right]+\frac{\tilde{g}_1^2}{18}
\left[\alpha^2-\frac{2}{3}\alpha-4+\frac{\pi^2}{4}
+i\pi\left(\alpha-\frac{1}{3}\right)\right]
\\
+ \frac{\tilde{g}_2}{6}
\left[\alpha^2+\frac{2}{3}\alpha-\frac{\pi^2}{12}
+i\pi\left(\alpha+\frac{1}{3}\right)\right] \end{array}\right\}
\frac{1}{\tilde{x}}
\\ \nonumber
&& \mbox{} - \frac{i}{3}\left\{
\tilde{g}_1+\frac{\tilde{g}_1^2}{3}
\left[\alpha-\frac{1}{3}+\frac{i\pi}{2}\right]+\frac{\tilde{g}_2}{3}
\right\} \frac{\ln \tilde{x}}{\tilde{x}}
\\ \nonumber
&& + \frac{i}{6}\left\{ \frac{\tilde{g}_1^2}{3}-\tilde{g}_2
\right\} \frac{(\ln \tilde{x})^2}{\tilde{x}}.
\end{eqnarray}
The exact asymptotic form for $y$ in the limit
$\tilde{x}\rightarrow 0$ is given by Eq.~(\ref{z-sol}). Expanding
this perturbatively as in Eq.~(\ref{zexp}) for small
$\xi\ln(1/\tilde{x})$, $\mbox{i.e.}$ for $\tilde{x}$ in the range
$1 \gg \tilde{x} \gg \exp(-1/\xi)$, gives the asymptotic form for
$y$ up to second-order corrections
\begin{equation}\label{asymp}
y(\tilde{x}) \rightarrow \sqrt{2k}\,\tilde{A}_k \tilde{f}_0
\frac{1}{\tilde{x}} + \sqrt{2k}\,\tilde{A}_k \tilde{f}_1 \
\frac{\ln \tilde{x}}{\tilde{x}}+\frac{1}{2}\sqrt{2k}\,\tilde{A}_k
\tilde{f}_2 \ \frac{(\ln \tilde{x})^2}{\tilde{x}}.
\end{equation}
Comparing this with Eq.~(\ref{final-y}), the coefficient of
$\tilde{x}^{-1}$ is the desired result because it will give
$\tilde{A}_k$ up to second-order corrections. The coefficient of
$\ln \tilde{x}/\tilde{x}$ simply give the consistent asymptotic
behavior, that is, proportional to $z_k$. Substituting
Eq.(\ref{g1}) into Eq.~(\ref{final-y}),
 matching the coefficient of $\tilde{x}^{-1}$ with that in
 Eq.(\ref{asymp}),
  the noncommutative power spectrum up to second order is
\begin{eqnarray}\label{amp}
\tilde{P}^{2nd}_{R_c}(k) &=&\frac{k^3}{2\pi^2}|\tilde{A}_k|^2
\\ \nonumber
&=&\frac{k^2}{(2\pi)^2}\frac{1}{\tilde{f}_0^2} \left[1 -
2\alpha\frac{\tilde{f}_1}{\tilde{f}_0}+\left(
3\alpha^2-4+\frac{5\pi^2}{12}\right)\left(\frac{\tilde{f}_1}{\tilde{f}_0}
\right)^2+\left(-\alpha^2+\frac{\pi^2}{12}
\right)\frac{\tilde{f}_2}{\tilde{f}_0}\right].
\end{eqnarray}
Substituting Eqs.(\ref{1/f0}), (\ref{f1/f0}) and  (\ref{f2/f0})
into Eq.(\ref{amp}) leads to
\begin{eqnarray}
\tilde{P}^{2nd}_{R_c}(k) =
          & & P^{2nd}_{R_c}(k)\\
          & & -  \fr{\mu_0}{2\epsilon_1}\frac{H^2}{(2\pi)^2}
              \left\{\begin{array}{l}
                  4+(32 \alpha - 10)\epsilon_1 + (8\alpha + 2)\delta_1 \\
                  + (96 \alpha^2 - 8\alpha - 232 + 24\pi^2)\epsilon_1^2
                  + (28\alpha^2+32\alpha-158+19\pi^2)\epsilon_1\delta_1 \\
                  + (12\alpha^2+6\alpha -16+{5\pi^2 \over 3})\delta_1^2
                       -(4\alpha^2-{\pi^2 \over 3})\delta_2
                     \end{array}
             \right\} \nonumber
\end{eqnarray}
where the commutative contribution is given by
\begin{eqnarray}
 \label{2ndps}
P^{2nd}_{R_c}(k) & = &  \fr{1}{2\epsilon_1}\frac{H^2}{(2\pi)^2}
\left\{ 1 -2\epsilon_1  + 2\alpha(2\epsilon_1+\delta_1)+
\left(4\alpha^2-23+\frac{7\pi^2}{3}\right)\epsilon_1^2 \right.\\
&& \left. +
\left(3\alpha^2+2\alpha-22+\frac{29\pi^2}{12}\right)\epsilon_1\delta_1
+ \left(3\alpha^2-4+\frac{5\pi^2}{12}\right)\delta_1^2 +
\left(-\alpha^2+\frac{\pi^2}{12}\right)\delta_2 \right\} \nonumber
\end{eqnarray}
and the right hand side should be evaluated at $k=aH$. Comparing
with the result Eq.(\ref{nonps11}) from the slow-roll
approximation, we find additional terms depending $\mu_0$ even for
up to first-order corrections. We identify these with  an effect
of choosing the first in Eq.(\ref{slow-di}). Making use of
Eq.(\ref{slow-di}), we  obtain the following
relations\footnote{Here we note that $\fr{\partial \mu_0}{\partial
t}=-2H\mu_0(1+\epsilon_1)$ and $\fr{\partial \mu_0}{\partial \ln
k}=2\mu_0$\cite{KLM}. Actually $\dot \mu_0=-4H\mu_0\epsilon_1$ in
Eq.(\ref{slow-di}) comes from taking into account these terms.}
for calculating the spectral index and running spectral index:
\begin{eqnarray}
&& \fr{d\mu_0}{d\ln
k}=\fr{1}{(1-\epsilon_1+4\mu_0\epsilon_1)H}\fr{\partial
\mu_0}{\partial t}+\fr{\partial \mu_0}{\partial \ln k}\simeq
-4\mu_0\epsilon_1,\\
&& \fr{d\epsilon_1}{d\ln k}=
\fr{1}{(1-\epsilon_1+4\mu_0\epsilon_1)H} \fr{\partial
\epsilon_1}{\partial t}\simeq 2(\epsilon_1^2+\epsilon_1\delta_1),\\
\nonumber &&\fr{d\delta_1}{d\ln
k}=\fr{1}{(1-\epsilon_1+4\mu_0\epsilon_1)H}\fr{\partial
\delta_1}{\partial t}\simeq
\epsilon_1\delta_1-\delta_1^2+\delta_2,
\\ \nonumber
&& \fr{d\delta_2}{d\ln
k}=\fr{1}{(1-\epsilon_1+4\mu_0\epsilon_1)H}\fr{\partial
\delta_2}{\partial t} \simeq
2\epsilon_1\delta_2-\delta_1\delta_2+\delta_3,\\
\nonumber &&\fr{d\delta_3}{d\ln
k}=\fr{1}{(1-\epsilon_1+4\mu_0\epsilon_1)H}\fr{\partial
\delta_3}{\partial t} \simeq
3\epsilon_1\delta_3-\delta_1\delta_3+\delta_4.
\end{eqnarray}
Then the  spectral index defined by
\begin{equation}
\tilde{n}_{s}(k) = 1 + \frac{d \ln \tilde{P}^{2nd}_{R_c}}{d \ln k}
\end{equation}
can be easily calculated up to third order
\begin{eqnarray}
\tilde{n}_{s}(k) = & &n_{s}(k)+\mu_0 \left\{
\begin{array}{l}
                       16\epsilon_1+
                      (32\alpha+12)\epsilon_1^2
                      - (32\alpha-10)\epsilon_1\delta_1
                      + 2\delta_1^2 -
                      2\delta_2 \\
                      +(32\alpha+12)\epsilon_1^3
                      +
                      \left(
                           -80\alpha^2-140\alpha+520-{148\pi^2\over 3}
                       \right)\epsilon_1^2\delta_1 \\
                      +\left(
                           16\alpha^2-16\alpha+68-{28\pi^2\over 3}
                       \right)\epsilon_1\delta_1^2 \\
                       +
                       \left(
                           16\alpha^2-20\alpha+64-{28\pi^2\over 3}
                       \right)\epsilon_1\delta_2
                      +4\alpha\delta_1^3-4\alpha\delta_1\delta_2
\end{array}
                      \right\}.
\end{eqnarray}
where the right hand side should be evaluated at $k=aH$.  The
commutative contribution up to third order is given by
\begin{eqnarray}\label{nonsin}
n_{s}(k) =& & 1 - 4\epsilon_1 - 2\delta_1 +(8\alpha-8)\epsilon_1^2
              + (10\alpha -6)\epsilon_1\delta_1 -2\alpha\delta_1^2+2\alpha\delta_2
              \\ \nonumber
          & & + \left(
                    -16\alpha^2+40\alpha-108+{28\pi^2\over 3}
                \right)\epsilon_1^3
                +\left(
                    -31\alpha^2+60\alpha-172+{199\pi^2\over 12}
                \right)\epsilon_1^2\delta_1
                \\ \nonumber
          & & + \left(
                    -3\alpha^2+4\alpha-30+{13\pi^2\over 4}
                \right)\epsilon_1\delta_1^2
                +\left(
                    -7\alpha^2+8\alpha-22+{31\pi^2\over 12}
                \right)\epsilon_1\delta_2
                \\ \nonumber
          & & + \left(
                    -2\alpha^2+8-{5\pi^2\over 6}
                \right)\delta_1^3
                +\left(
                    3\alpha^2-8+{3\pi^2\over 4}
                \right)\delta_1\delta_2
                +\left(
                    -\alpha^2+{\pi^2\over 12}
                \right)\delta_3.
\end{eqnarray}

 Finally the running
spectral index up to third order is given by
\begin{eqnarray}
\frac{d\tilde{n}_s}{d\ln k} =& &\frac{d n_s}{d\ln k}+ \mu_0\left\{
\begin{array}{l}
-32\epsilon_1^2+32\epsilon_1\delta_1-32\epsilon_1^3+(160\alpha+70)\epsilon_1^2\delta_1
+(-32\alpha+6)\epsilon_1\delta_1^2 \\
+(-32\alpha+14)\epsilon_1\delta_2-4\delta_1^3+6\delta_1\delta_2-2\delta_3 \\
+(64\alpha -8)\epsilon_1^4
-\left(80\alpha^2-212\alpha-662+\frac{148\pi^2}{3}\right)\epsilon_1^3\delta_1
\\
-\left(240\alpha^2+452\alpha-1566+148\pi^2\right)\epsilon_1^2\delta_1^2
\\
-\left(80\alpha^2+172\alpha-534+\frac{148\pi^2}{3}\right)\epsilon_1^2\delta_2
\\
 -(4\alpha+4)\epsilon_1\delta_1^3
+\left(48\alpha^2-48\alpha+206-28\pi^2\right)\epsilon_1\delta_1\delta_2
\\
+\left(16\alpha^2-20\alpha+62-\frac{28\pi^2}{3}\right)\epsilon_1\delta_3
\\
-12\alpha\delta_1^4+20\alpha\delta_1^2\delta_2-4\alpha\delta_2^2-4\alpha\delta_1\delta_3
\end{array}
 \right\}.
\end{eqnarray}
Also the commutative contributions up to third order take the form
\begin{eqnarray} \label{nonrsin}
\frac{d}{d\ln k} n_{s} & = &
-8\epsilon^{2}_{1}-10\epsilon_1\delta_1+2\delta^{2}_{1}-2\delta_2+(32\alpha
-40)\epsilon^{3}_{1}+(62\alpha -60)\epsilon^{2}_{1}\delta_1
\\ \nonumber
&+&(6\alpha-4)\epsilon_1\delta^{2}_{1}+(14\alpha
-8)\epsilon_1\delta_2+4\alpha\delta^{3}_{1}-6\alpha\delta_1\delta_2+2\alpha\delta_3 \\
\nonumber
         &+&\left(-96\alpha^2+272\alpha-688+56\pi^2\right)\epsilon_1^4
            +\left(-251\alpha^2+602\alpha-1568+{1667\pi^2\over
             12}\right)\epsilon_1^3\delta_1
               \\ \nonumber
         &+&\left(-105\alpha^2+202\alpha-640+{251\pi^2\over 4}
             \right)\epsilon_1^2\delta_1^2
            +\left(-59\alpha^2+106\alpha-268+{323\pi^2\over
             12}\right)\epsilon_1^2\delta_2
               \\ \nonumber
         &+&\left(-6\alpha^2+4\alpha+24-{5\pi^2\over
             2}\right)\epsilon_1\delta_1^3
            +\left(-4\alpha^2+10\alpha-106+{34\pi^2\over
             3}\right)\epsilon_1\delta_1\delta_2
               \\ \nonumber
         &+&\left(-10\alpha^2+10\alpha-22+{17\pi^2\over
             6}\right)\epsilon_1\delta_3
            +\left(6\alpha^2-24+{5\pi^2\over 2}\right)\delta_1^4
               \\ \nonumber
         &+&\left(-12\alpha^2+40-4\pi^2\right)\delta_1^2\delta_2
            +\left(4\alpha^2-8+{2\pi^2\over 3}\right)\delta_1\delta_3
               \\ \nonumber
         &+&\left(3\alpha^2-8+{3\pi^2\over 4}\right)\delta_2^2
            +\left(-\alpha^2+{\pi^2\over 12}\right)\delta_4.
            \nonumber
\end{eqnarray}

Up to now our calculation was  done at the pivot scale of $k=aH$.
In order to obtain the correct expressions for noncommutative
spacetime inflation, we have to change the pivot scale: $k=aH\to
k=\sqrt{2}aH$. Up to second-order corrections, we assume that
$k=aH\to k=\sqrt{2}aH$ corresponds to $\alpha=0.7296 \to
\alpha_*=0.3831$ in the above expressions\cite{STW}. Further,
assuming that $\triangle
\tilde{\eta}=\tilde{\eta_*}-\tilde{\eta_c}$ is very small, we
neglect higher order corrections like $(\triangle \tilde{\eta})^n$
for $n\ge2$ in obtaining the noncommutative power spectrum at
$\tilde{\eta}=\tilde{\eta}_*$ from the noncommutative power
spectrum at $\tilde{\eta}=\tilde{\eta}_c$.

  As an example, we choose the
power-law inflation like $a(t)\sim t^p$ whose potential is given
by \beq V(\phi)=V_0 \exp\Big(-\sqrt{\fr{2}{p}}\phi\Big).\eeq Thus
slow-roll parameters are determined  by \beq \label{pislp}
\epsilon_1=\fr{1}{p},~\delta_1=-\fr{1}{p},~
\delta_2=2\delta_1^2=\fr{2}{p^2},~\delta_3=6\delta_1^3=-\fr{6}{p^3}
,~\delta_4=24\delta_1^4=\fr{24}{p^4}.\eeq Then the noncommutative
power spectrum takes the form \beq
\label{pips1}\tilde{P}^{PI,2nd}_{R_c}(k)  =
P^{PI,2nd}_{R_c}(k)+\frac{\mu_0H^4}{(2\pi)^2\dot\phi^2} \left\{
       -4  +\fr{12(1-2\alpha_*)}{p}+\fr{1}{p^2}\Big(90+34\alpha_*
      -72\alpha_*^2 -\fr{22\pi^2}{3} \Big)
\right\}, \eeq where the numerical values of the coefficients in
this equation are $-4,2.8056,20.0812$, respectively.

 Here $\mu_0$-independent terms are given by
 \beq
 \label{2ndpss}
P^{PI,2nd}_{R_c}(k) = \frac{H^4}{(2\pi)^2\dot\phi^2} \left\{ 1
+\fr{2(\alpha_*-1)}{p}+\fr{1}{p^2}\Big(2\alpha_*^2-2\alpha_*-5
       +\fr{\pi^2}{2} \Big)\right\}, \nonumber \eeq
where the numerical values of the coefficients  are $1,-1.2338,
-0.537867$, respectively. The noncommutative spectral index can be
easily calculated up to third order
\begin{equation} \label{pisi1}
\tilde{n}^{PI}_{s}(k) = n^{PI}_{s}(k)+\mu_0 \left\{\fr{16}{p} +
\fr{64\alpha_*}{p^2}+\fr{1}{p^3}\Big(128\alpha_*^2+120\alpha_*-312+64\pi^2\Big)\right\},
\end{equation}
where the numerical values of the coefficients in this expression
are 16, 24.5184, 384.413, respectively. Here \beq \label{pisi11}
n^{PI}_{s}(k)=1-\fr{2}{p}-\fr{2}{p^2}-\fr{2}{p^3}. \eeq Finally
the running spectral index is found to be \beq
\label{pirsi1}\frac{d\tilde{n}^{PI}_s}{d\ln k} = \frac{d
n^{PI}_s}{d\ln k} -\mu_0 \left\{\fr{64}{p^2}
+\fr{8(32\alpha_*+8)}{p^3}+\fr{1}{p^4}\Big(512\alpha_*^2+736\alpha_*-1184
+\fr{256\pi^2}{3} \Big)\right\},\eeq where the numerical values of
the coefficients in this equation are 64, 162.074, 15.3118,
respectively. Here the commutative contribution is zero up to
fourth order, \beq \label{pirsi11}\frac{d n^{PI}_s}{d\ln k}=0.
\eeq Comparing the above expressions with those of
$\mu_0$=constant in Ref.\cite{HM3}, we find additional corrections
in equations (\ref{pips1}), (\ref{pisi1}), and (\ref{pirsi1}).
These are replacement of $\alpha \to \alpha_*$ from the change of
pivot scale and additional contributions from a noncommutative
parameter with $\mu_0\not=$ constant. Actually the results from
the slow-roll approximation with $\mu_0$=constant are those in the
first terms $-4\mu_0, \fr{16\mu_0}{p}, -\fr{64\mu_0}{p^2}$ in Eqs.
(\ref{pips1}), (\ref{pisi1}), and (\ref{pirsi1}), respectively. On
the other hand, the slow-roll expansion with $\mu_0 \not=$
constant gives us $\tilde{P}_{R_c} \propto
\mu_0\Big(-4+\fr{2.8056}{p}+\fr{20.0812}{p^2}\Big)$, $\tilde{n_s}
\propto
\mu_0\Big(\fr{16}{p}+\fr{24.5184}{p^2}+\fr{384.413}{p^3}\Big)$,
$\fr{d\tilde{n}_s}{d \ln k} \propto
-\mu_0\Big(\fr{64}{p^2}+\fr{162.074}{p^3}+\fr{15.3118}{p^4})$.
Hence the slow-roll expansion decreases less the power spectrum
than that of slow-roll approximation. On the other hand, the
slow-roll expansion increases more the spectral index than that of
slow-roll approximation. Finally, the slow-roll expansion
decreases more the running spectral index than that of slow-roll
approximation.

\section{Slow-roll expansion with $\mu_0$=constant}
In the previous section we consider $\mu_0(t,k)$ to be a function
of time $t$ and the comoving Fourier modes $k$ in the beginning.
It is noted that near the horizon crossing time ($k\propto aH$),
one has $\mu_0 \propto H^4$. This implies that $\mu_0$ may be
considered to be nearly constant. In this section we use the case
of $\mu_0$=constant to calculate the power spectrum and
$\fr{\mu_0}{d\ln k}\simeq -4\mu_0\epsilon_1$ to compute the
spectral index and running spectral index. We call it the
slow-roll expansion with $\mu_0$=constant. This case provides us a
new result  between the slow-roll approximation in Sec. II and the
slow-roll expansion in Sec. III.

Following the previous section, the noncommutative power spectrum
is calculated as

\begin{eqnarray}
\tilde{P}^{2nd}_{R_c}(k) =
          & & P^{2nd}_{R_c}(k)\\
          & & -  \fr{\mu_0}{2\epsilon_1}\frac{H^2}{(2\pi)^2}
              \left\{\begin{array}{l}
                  4+(16 \alpha - 2)\epsilon_1 + (8\alpha + 2)\delta_1 \\
                  + (16 \alpha^2 + 28\alpha - 84 + {28\pi^2\over 3})\epsilon_1^2
                  + (12\alpha^2+32\alpha-78+{29\pi^2\over 3})\epsilon_1\delta_1 \\
                  + (12\alpha^2+6\alpha -16+{5\pi^2 \over 3})\delta_1^2
                       -(4\alpha^2-{\pi^2 \over 3})\delta_2
                     \end{array}
             \right\}. \nonumber
\end{eqnarray}

The spectral index takes the form

\begin{eqnarray}
\tilde{n}_{s}(k) = & &n_{s}(k)+\mu_0 \left\{
\begin{array}{l}
                      16\epsilon_1
                      +28\epsilon_1^2
                      -6\epsilon_1\delta_1
                      +2\delta_1^2
                      -2\delta_2 \\
                      +28\epsilon_1^3
                      -\left( 12\alpha+72\right)\epsilon_1^2\delta_1
                      -12\epsilon_1\delta_1^2 \\
                      -\left(4\alpha+16\right)\epsilon_1\delta_2
                      +4\alpha\delta_1^3-4\alpha\delta_1\delta_2
\end{array}
                      \right\}.
\end{eqnarray}

Finally the running spectral index is

\begin{eqnarray}
\frac{d\tilde{n}_s}{d\ln k} =& &\frac{d n_s}{d\ln k}+\mu_0\left\{
\begin{array}{l}
-32\epsilon_1^2+32\epsilon_1\delta_1
-32\epsilon_1^3+150\epsilon_1^2\delta_1
-10\epsilon_1\delta_1^2-2\epsilon_1\delta_2 \\
-4\delta_1^3+6\delta_1\delta_2-2\delta_3+24\epsilon_1^4
-(12\alpha-246)\epsilon_1^3\delta_1
\\
-(36\alpha+226)\epsilon_1^2\delta_1^2
-(12\alpha+74)\epsilon_1^2\delta_2
\\
-(4\alpha+4)\epsilon_1\delta_1^3 -34\epsilon_1\delta_1\delta_2
-(4\alpha+18)\epsilon_1\delta_3
\\
-12\alpha\delta_1^4+20\alpha\delta_1^2\delta_2
-4\alpha\delta_2^2-4\alpha\delta_1\delta_3
\end{array}
 \right\}.
\end{eqnarray}

Choosing the power-law inflation, one has explicit forms for the
power spectrum, spectral index, and running spectral index. The
power spectrum is \beq \label{pips2}\tilde{P}^{PI,2nd}_{R_c}(k) =
P^{PI,2nd}_{R_c}(k)+\frac{\mu_0H^4}{(2\pi)^2\dot\phi^2} \left\{
       -4
       +\fr{4(1-2\alpha_*)}{p}+\fr{1}{p^2}\Big(22-8\alpha_*^2-2\alpha_*
       -2\pi^2 \Big)
\right\}. \eeq

The spectral index is given by
\begin{equation} \label{pisi2}
\tilde{n}^{PI}_{s}(k) = n^{PI}_{s}(k)+\mu_0 \left\{\fr{16}{p} +
\fr{32}{p^2}+\fr{1}{p^3}\Big(8\alpha_*+56\Big)\right\}.
\end{equation}

The running spectral index  takes the form as \beq
\label{pirsi2}\frac{d\tilde{n}^{PI}_s}{d\ln k}= \frac{d
n^{PI}_s}{d\ln k} -\mu_0 \left\{\fr{64}{p^2}
+\fr{192}{p^3}+\fr{1}{p^4}\Big(32\alpha_*+416\Big)\right\}. \eeq
 Finally, the slow-roll
expansion with $\mu_0=$ constant gives us $\tilde{P}_{R_c} \propto
\mu_0\Big(-4+\fr{0.9352}{p}+\fr{0.3205}{p^2}\Big)$, $\tilde{n_s}
\propto \mu_0\Big(\fr{16}{p}+\fr{32}{p^2}+\fr{59.0648}{p^3}\Big)$,
$\fr{d\tilde{n}_s}{d \ln k} \propto
-\mu_0\Big(\fr{64}{p^2}+\fr{192}{p^3}+\fr{428.257}{p^4})$. Hence
the slow-roll expansion with $\mu_0$=constant decreases more the
power spectrum than that of slow-roll expansion with
$\mu_0\not=$constant. On the other hand, the slow-roll expansion
with $\mu_0$=constant increases more the spectral index up to
second order than that of slow-roll expansion with
$\mu_0\not=$constant. Finally, the slow-roll expansion with
$\mu_0$=constant decreases more the running spectral index up to
third order than that of slow-roll expansion with
$\mu_0\not=$constant.

\section{Discussion}

We investigate an effect of the uncertainty of noncommutative
spacetime  on the period of slow-roll inflation. For our purpose,
the noncommutative parameter $\mu_0(t,k)=(kH/aM_s^2)^2$ arisen
from the time uncertainty of $\triangle \tau= k/M_s^2$ is taken to
be a zeroth order of slow-roll parameter. This is a key step in
calculation of  the noncommutative  power spectrum. Actually
$\mu_0$ is very small and thus one recovers the commutative result
when $M_s \to \infty$. When comparing with other slow-roll
parameters $\epsilon_1(t)$ and $\delta_n(t)$, a difference is that
$\mu_0$ is a function of time $t$ as well as comoving Fourier mode
$k$.  We calculate the noncommutative power spectrum up to second
order in the slow-roll expansion. In deriving it,  we have to
remind the reader two important facts. First the slow-roll
expansion is originally designed to  calculate the power spectrum
of curvature perturbation at the horizon crossing time when
$k_c=a(\tilde{\eta_c})H(\tilde{\eta_c})$. However, in the
noncommutative inflation, the horizon crossing time is delayed as
$k_*=\sqrt{2}a(\tilde{\eta_*})H(\tilde{\eta_*})$. In the
first-order corrections, this change of the pivot scale amount to
replacing $\alpha=2-\ln2-\gamma=0.7296$ by
$\alpha_*=\alpha-\ln2/2=0.3831$. Although we don't know whether or
not this substitution rule  is valid up to second-order
corrections,  we assume this rule to be valid for  calculating the
noncommutative power spectrum up to second order.

Second we assume that $\triangle
\tilde{\eta}=\tilde{\eta_*}-\tilde{\eta_c}$ is very small, thus we
neglect higher order corrections like $(\triangle \tilde{\eta})^n$
for $n\ge2$ in calculating the noncommutative  power spectrum at
$\tilde{\eta}=\tilde{\eta}_*$ from the noncommutative power
spectrum at $\tilde{\eta}=\tilde{\eta}_c$. This implies that the
form of the power spectrum  at $\tilde{\eta}=\tilde{\eta}_c$
persists for the time at $\tilde{\eta}=\tilde{\eta}_*$.

Further we use the case of $\mu_0$=constant to calculate the power
spectrum and $\fr{\mu_0}{d\ln k}\simeq -4\mu_0\epsilon_1$ to
compute the spectral index and running spectral index.  This case
provides us an intermediate result  between the slow-roll
approximation with $\mu_0$=constant and the slow-roll expansion
with $\mu_0\not=$constant.

In conclusion, we observe important facts  from our calculation
using the slow-roll expansion. We find corrections arisen from a
change of the pivot scale
 and  the presence of a variable noncommutative parameter,
 when comparing with the commutative power spectrum. The
power-law inflation is chosen to obtain  explicit forms for the
power spectrum, spectral index, and running spectral index. We
find from Eqs.(\ref{pips1}) and (\ref{2ndpss}) that the
noncommutative power spectrum always appears to be smaller than
that of the commutative case. The noncommutative effect of
higher-order corrections compensates for a loss of higher-order
corrections in the commutative form.  This suppression propagates
the spectral index in a way that it is larger than the one in the
commutative case. As is shown in Eqs.(\ref{pisi1}) and
(\ref{pisi11}), the noncommutative effect of higher-order
corrections again compensates for a loss of higher-order
corrections in the commutative case. Furthermore we note that the
spectral index for commutative form is invariant under transform
of the pivot scale.  According to the WMAP data\cite{Wmap1}, one
finds $\fr{d n_s}{d \ln k}=-0.031^{+0.016}_{-0.017}$ at
$k$=0.05Mpc$^{-1}$. From Eq.(\ref{pirsi11}), it is obvious that
there is no contribution to the running spectral index from the
commutative power-law inflation. It seems that a way to explain
the above WMAP date is to consider the spacetime uncertainty
during inflation. As is shown in Eq.(\ref{pirsi1}),  all
higher-order corrections to the commutative case always provide us
negative spectral indexes. Finally, we find the same observations
even for the slow-roll expansion with $\mu_0$=constant.
\subsection*{Acknowledgements}
Y.S.  was supported in part by  KOSEF, Project No.
R02-2002-000-00028-0. H.W. was in part supported by KOSEF,
Astrophysical Research Center for the Structure and Evolution of
the Cosmos.


\begin{thebibliography}{99}
\bibitem{Bran1} R. Brandenberger, hep-th/0103156.

\bibitem{Yone} T. Yoneya, in {\it Wandering in the Fields}, eds. K.
Kawarabayashi and  A. Ukawa (World Scientific, 1987), p. 419; M.
Li
 and T. Yoneya, Phys. Rev. Lett. {\bf 78} (1977)
 1219 [hep-th/9611072];
 T. Yoneya, Prog. Theor. Phys. {\bf 103} (2000)
 1081 [hep-th/0004074];
 J. Polchinski, {\it String Theory} (Cambridge Univ. Press,
 Cambridge, 1998), vol. II.

\bibitem{BH} R. Brandenberger and P. M. Ho, Phys. Rev. D {\bf 66}
(2002) 023517 [hep-th/0203119].

\bibitem{Wmap1} H. V. Peiris, {\it et al}, Astrophys. J. Suppl. {\bf 148} (2003) 213
[astro-ph/0302225];
C. L. Bennett, {\it et al}, Astrophys. J. Suppl. {\bf 148} (2003)
1 [astro-ph/0302207];
D. N. Spergel, {\it et al}, Astrophys. J. Suppl. {\bf 148} (2003)
175 [astro-ph/0302209].

\bibitem{Wmap2} P. Mukherjee and Y. Wang, Astrophys. J. {\bf 599} (2003) 1
[astro-ph/0303211];
S. L. Bridle, A. M. Lewis, J. Weller and G. Efstathiou, Mon. Not.
Roy. Astron. Soc. {\bf 342} (2003) L72 [astro-ph/0302306].

\bibitem{LMMP}  F. Lizzi, G. Mangano, G. Miele, and  M. Peloso,
 JHEP {\bf 0206} (2002) 049 [hep-th/0203099].


\bibitem{HM1} Q. G. Huang and M. Li, JHEP {\bf 0306} (2003)
014 [hep-th/0304203].

\bibitem{TMB} S. Tsujikawa, R. Maartens, and R. Brandenberger, Phys. Lett. B {\bf 574} (2003) 141
[astro-ph/0308169].

\bibitem{HM2} Q. G. Huang and M. Li, JCAP {\bf 0311} (2003)
001 [astro-ph/0308458].

\bibitem{HM3} Q. G. Huang and M. Li, astro-ph/0311378.

\bibitem{KLM} Hungsoo Kim, G. S. Lee, and Y. S. Myung,
hep-th/0402018.

\bibitem{bardeen} J. M. Bardeen, Phys. Rev. D{\bf 22} (1980) 1882;
H. Kodama and M. Sasaki, Prog. Theor. Phys. Supp. {\bf 78}
(1984)1.

\bibitem{lukash} V. N. Lukash, JETP Lett. {\bf 31} (1980) 596;
V. N. Lukash, Sov. Phys. JETP {\bf 52} (1980) 807.

\bibitem{mukhanov} V. F. Mukhanov, JETP Lett. {\bf 41} (1985) 493.

\bibitem{LL} A. R. Liddle and D. H. Lyth, {\it Cosmological
inflation and Large-Scale Structure} (Cambridge Univ. Press,
Cambridge, 2000), p. 184.

\bibitem{BG} O. Bertolami and  L. Guisado,
Phys. Rev. D{\bf 67 } (2003) 025001 [gr-qc/0207124];
{\it ibid}, JHEP {\bf 0312} (2003) 013 [hep-th/0306176].


\bibitem{FKM} M. Fukuma, Y. Kono, and A. Miwa, hep-th/0307029;
{\it ibid}, hep-th/0312298;
{\it ibid}, hep-th/0401153.

\bibitem{SL} E. D. Stewart and D. H. Lyth, Phys. Lett. B{\bf 302} (1993) 171.


\bibitem{MS1}
J. Martin and D. J. Schwarz, Phys. Rev. D{\bf 62} (2000) 103520
[astro-ph/9911225];
J. Martin , A. Liazuelo, and D. J. Schwarz, Astrophys. J. {\bf
543} (2000)L99 [astro-ph/0006392];

\bibitem{STW} E. Stewart, Phys. Rev. D {\bf 65} (2002) 103508
 [astro-ph/0110322].

\bibitem{MS2}
J. Martin and D. J. Schwarz, Phys. Rev. D {\bf 67} (2003) 083512
[astro-ph/0210090];

\bibitem{SG} E. Stewart and J.-O. Gong, Phys. Lett. B {\bf 510} (2001) 1
[astro-ph/0101225].

\end{thebibliography}
\end{document}